\documentclass[a4paper,12pt]{article}
\usepackage[english]{babel}
\usepackage[numbers,comma,square,compress]{natbib}
\usepackage{amssymb,amsmath,amsfonts,bm,enumerate}
\usepackage{graphicx}% Include figure files

\usepackage{color}
\definecolor{hlcolor}{rgb}{0.5,0,0.8}

\usepackage{ifpdf}
\ifpdf  % Compiling PDF with pdflatex
 \usepackage[pdftex,a4paper,hmargin=25mm,vmargin=25mm]{geometry}
 \usepackage[pdftex,colorlinks=true,allcolors=blue]{hyperref}
\else  % Compiling DVI with latex
 \usepackage[dvips,a4paper,hmargin=25mm,vmargin=25mm]{geometry}
 \usepackage[dvips,colorlinks=true,allcolors=blue]{hyperref}
 \fi

\DeclareMathAlphabet{\mathcal}{OMS}{cmsy}{b}{n}

\newcommand{\email}[1]{\footnote{E-mail: \href{mailto:#1}{#1}}}

\begin{document}

\title{  Tree-level processes in very special relativity}
\author{R. Bufalo\email{rodrigo.bufalo@ufla.br} and T. Cardoso e Bufalo\email{tati.cardoso@ufla.br} \\
%EndAName
\text{$^{1}$ \small Departamento de F\'isica, Universidade Federal de Lavras,}\\
\text{ \small Caixa Postal 3037, 37200-900 Lavras, MG, Brazil}\\
\\
}

\maketitle
\date{}

\begin{abstract}
In this paper, we discuss the Bhabha and Compton scattering for the quantum electrodynamics defined in the framework of SIM(2) very special relativity (VSR).
The main aspect of the VSR setting is that it admits different types of interactions appearing in a nonlocal form due to the modified gauge invariance.
We explore the richness of these new couplings in the evaluation of the differential cross section for these tree-level processes.
We assess the behavior of the leading VSR Lorentz violation modifications by considering some special limits for the Bhabha and Compton cross section expressions.
\end{abstract}

\section{Introduction}

In the description of nature, our experimentally verified theories, such as the Standard Model (SM) of particles and General Relativity, are fundamentally understood in terms of gauge invariance and spacetime symmetries.
Due to its importance in the formulation of the given field theories,  the validity of these exact symmetries has been scrutinized by a number of precision tests \cite{ref53,Jacobson:2005bg,AmelinoCamelia:2008qg}.
In one hand, any deviation from these symmetries is expected to signal manifestations of physics beyond the Standard Model, that are hoped to be revealed in high-energy experiments.
On the other hand, physical phenomena that are not adequately explained by the known theories, e.g., neutrino masses, matter-antimatter asymmetry, quantum gravity, etc. \cite{Lykken:2010mc}, can be suitably described by the addition of new degrees of freedom (d.o.f.), that ultimately are governed by a new symmetry principle.

In recent years, we have seen a great amount of interest in field theories formulated in a Lorentz violating framework \cite{ref53}.
In particular, the most interesting proposals are those that preserve the basic elements of special relativity, because they are in agreement with well-understood physics, but additionally these modified models allow the description of new and unexplored phenomena.
Within this context, a framework satisfying the above criteria is the  Cohen and Glashow very special relativity (VSR) \cite{Cohen:2006ky,Cohen:2006ir}.
The main aspect in the VSR proposal is that the laws of physics are invariant under the subgroups of the Poincar\'{e} group preserving the basic elements of special relativity.
In the VSR framework, we found a phenomenologically rich scenario because it presents a modified gauge symmetry, admitting a variety of new gauge invariant interactions.
Many interesting theoretical and phenomenological aspects of VSR effects have
been extensively discussed \cite{Dunn:2006xk,Alfaro:2015fha,Lee:2015tcc,Nayak:2016zed,Alfaro:2017umk,Alfaro:2019koq,Bufalo:2019qot}.

In relation to the kinematics of the VSR framework, there are two subgroups satisfying the prior requirements, namely, the HOM(2) (with three parameters) and the SIM(2) (with four parameters).
The former is the so-called Homothety group, generated by $T_1 = K_x + J_y$, $T_2 = K_y-J_x$, and $K_z$ ($\vec{J}$ and $\vec{K}$ are the generators of rotations and boosts, respectively).
The latter, called the similitude group SIM(2), is the HOM(2) group added by the $J_z$ generator.
Moreover, the symmetry groups SIM(2) and HOM(2) have the property of preserving the direction of a lightlike four-vector $n_{\mu}$ by scaling, transforming as $n \to e^{\varphi} n$ under boosts in the z direction.
This feature implies that theories invariant under one of these two subgroups have a preferred direction in the Minkowski spacetime, where Lorentz violating terms can be constructed as ratios of contractions of the vector $n_{\mu}$ with other kinematic vectors \cite{Cohen:2006ky}.

On the dynamical side of the VSR framework, we can make use of the last comment on how to write Lorentz violating but VSR invariant terms, to construct VSR covariant field theories.
As an illustration, one can simply write down a SIM(2) VSR-covariant Dirac equation in the form
\begin{equation}
\left(i\gamma^\mu \tilde{\partial}_\mu -m_e \right)\psi \left(x\right)=0,
\end{equation}
where the wiggle derivative operator is defined by $\tilde{\partial}_{\mu}=\partial_{\mu}+\frac{1}{2}\frac{m^{2}}{n.\partial}n_{\mu}$.
First, we see that the Lorentz violation appears in a nonlocal form, \footnote{ We will discuss below the subtle points involving the perturbative analysis of VSR nonlocal couplings.} the parameter $m$ sets the scale for the VSR effects, and the preferred null direction is chosen as $n_{\mu}=\left(1,0,0,1\right)$.

We can highlight an import feature of the VSR framework in field theories that is the addition of a gauge invariant massive mode, by presenting the SIM(2) VSR Maxwell equations $\tilde{\partial}_\mu F^{\mu \nu}=0$, where the field strength is defined in terms of the wiggled derivative as $\tilde{F}_{\mu\nu}=\tilde{\partial}_{\mu}A_{\nu}-\tilde{\partial}_{\nu}A_{\mu}$  \cite{Cheon:2009zx}.
These equations can be written in the VSR Lorenz condition $\tilde{\partial}_{\mu}A^{\mu}=0$, resulting in
\begin{equation}
\widetilde{\Box} A_\mu =0 \to (\Box + m^2)A_\mu =0
\end{equation}
showing that each component of the gauge field satisfies a massive Klein-Gordon equation.
This discussion is based on the fact that the Abelian gauge field has a VSR modified transformation law $\delta A_\mu = \tilde{\partial}_\mu \Lambda$.
In this case, the extended VSR gauge symmetry provides a suitable framework to describe massive modes without changing the number of physical polarization states of the photon \cite{Cheon:2009zx,Alfaro:2013uva}.
It is worth noting that massive modes of photons and their stability are a recurrent
subject of analysis in recent literature \cite{Heeck:2013cfa,Bonetti:2017pym}.

Among the most interesting works analyzing the phenomenological aspects of VSR effects, we can refer to the scattering of fermions by an external field and bremsstrahlung  in the QED \cite{Alfaro:2019koq}, one-loop corrections of QED \cite{Alfaro:2017umk}, pion and kaon decay \cite{Nayak:2016zed}, and the electroweak theory \cite{Dunn:2006xk,Alfaro:2015fha}.
However, since we have the presence of VSR nonlocal terms implying massive modes and new gauge invariant couplings, there are still further scattering processes involving electrons, such as Bhabha, M{\o}ller, and Compton, that could be investigated in this Lorentz violating framework in order to highlight the behavior of the VSR effects.
In this sense, we believe that the present analysis will complement the study of phenomenological aspects of QED within the framework of VSR.
Recently, many works have discussed electron scattering processes in order to elucidate the physical aspects of different Lorentz violating scenarios \cite{Altschul:2004xp,Kant:2009pm,Charneski:2012py,Casana:2012vu,Bufalo:2015eia,deBrito:2016zav,Santos:2018acs,Souza:2019zyt}.
Hence, the main purpose of the present work is the discussion and evaluation of the cross section for the Bhabha and Compton scattering processes defined in the VSR quantum electrodynamics.

In this paper, we examine the Lorentz violating VSR effects in the quantum electrodynamics in the framework of the Bhabha and Compton scattering.
In Sec.~\ref{sec2}, we establish the main aspects related to the SIM(2) VSR gauge invariance, presenting the dynamics and completeness relations for the fermion and gauge fields, necessary for the evaluation of the Feynman amplitudes in the VSR electrodynamics.
We compute in detail the VSR leading modifications of the cross section for the Bhabha and Compton scattering in Sec.~\ref{sec3}.
In order to highlight the VSR effects, we evaluate the Bhabha cross section in the high-energy limit.
Moreover, since the Compton energy shift, coming from energy-momentum conservation, is modified in the VSR framework, we have to revise and present a new expression for the differential cross section consistent with VSR.
The profile of the VSR modifications in the Compton scattering is discussed in the limit of low-energy and high-energy photons, where corrections are found for the Thomson cross section and Klein-Nishina formula.
In Sec.\ref{conc}, we summarize the results and present our final remarks.

%%%%%%%%%%%%%%%%%%%%%%%%%%%%%%%%%%%%
%%%%%%%%%%%%%%%%%%%%%%%%%%%%%%%%%%%%

\section{Gauge fields in VSR}
\label{sec2}
 
We start by considering a Lagrangian density for the gauge and SIM(2) VSR-invariant QED written 
as
\begin{equation}
\mathcal{L}=-\frac{1}{4}\tilde{F}_{\mu\nu}\tilde{F}^{\mu\nu}-\frac{1}{2\alpha}\left(\tilde{\partial}_{\mu}A^{\mu}\right)^{2}+\overline{\psi}\left[i\gamma^{\mu}\nabla_{\mu}-m_{e}\right]\psi \label{eq:3}
\end{equation}
where we have chosen the VSR modified Lorenz condition $\Omega\left[A\right]=\tilde{\partial}_{\mu}A^{\mu}=0$ and the field strength is defined in terms of the wiggled
derivative as $\tilde{F}_{\mu\nu}=\tilde{\partial}_{\mu}A_{\nu}-\tilde{\partial}_{\nu}A_{\mu}$. 
The VSR invariance expressed in terms of the wiggle derivative $\tilde{\partial}_{\mu}=\partial_{\mu}+\frac{1}{2}\frac{m^{2}}{n.\partial}n_{\mu}$ imposes a change in the gauge structure for the QED, in this sense the minimal coupling among the fermion and photon fields is determined by a new gauge invariant covariant derivative $\nabla_{\mu}$.
This new operator can be determined by making use of the SIM(2) gauge transformation $\delta A_{\mu}=\tilde{\partial}_{\mu}\Lambda$ and imposing the known transformation law $\delta\left(\nabla_{\mu}\psi\right)=i\Lambda\left(\nabla_{\mu}\psi\right)$ valid for any charged field $\psi$. Under these conditions, one can determine that the expression
\begin{equation}
\nabla_{\mu}\psi=D_{\mu}\psi 	+\frac{1}{2} \frac{m^2}{\left(n\cdot D\right)} n_{\mu}\psi, \label{eq:2a}
\end{equation}
satisfies the required properties and we have used the ordinary covariant derivative $D_\mu = \partial_\mu -ie A_\mu $.
Moreover, this definition reduces to the wiggle derivative $\tilde{\partial}_{\mu}$ in the noninteracting case. 

Some important remarks about the expression \eqref{eq:2a} are in place.
The nonlocal character of term $1/\left(n.D\right)$ in \eqref{eq:2a} implies the presence of an infinite number of interactions (in the coupling $e$).
The Feynman rules for these interactions can be obtained within the Wilson lines approach, which expresses the  respective terms in a suitable form with $N=1,2,3,...$ legs of photon fields \cite{Dunn:2006xk}, making the perturbative analysis workable.
In addition to the rule for the cubic $\left\langle \overline{\psi} \psi A \right\rangle $ vertex used in the computation of the Bhabha and Compton scattering processes, we must consider an additional Feynman rule for the Compton process in the  VSR setting, the quartic $\left\langle \overline{\psi} \psi A A\right\rangle $ vertex, with two photon external legs.
This nonlocal quartic vertex in the VSR electrodynamics is similar to the presence of the seagull diagram for the Compton scattering in the scalar QED.

The 1PI vertex function $\left\langle \overline{\psi} (p_1)\psi (p_2)A(p_3)\right\rangle $ can be obtained from the Lagrangian density \eqref{eq:3}, resulting in  \cite{Dunn:2006xk}
\begin{align} \label{eq.17}
\Lambda^{\mu}\left(p_{1},p_{2},p_{3}\right) & = -ie \left[\gamma^{\mu}+\frac{m^{2}}{2}
\frac{\left(\gamma.n\right)n^{\mu}}{\left(n.p_{1}\right)\left(n.p_{2}\right)}\right],
\end{align}
whereas the vertex function for the $\left\langle \overline{\psi} (p_1)\psi (p_2)A(p_3)A(p_4)\right\rangle $ reads  \cite{Dunn:2006xk}
\begin{align} \label{eq.17b}
\Xi^{\mu\nu}\left(p_{1},p_{2},p_{3},p_{4}\right) & = -\frac{ie^2 m^2}{2} \frac{(n.\gamma) n^\mu n^\nu }{(n.p_3)(n.p_4)} \left[  \frac{1}{\left(n.p_{1}\right)} + \frac{1}{\left(n.p_{2}\right)} - \frac{1}{n.\left(p_{1}+p_3\right)}- \frac{1}{n.\left(p_{1}+p_4\right)}   \right].
\end{align}
Observe that the VSR contribution to these vertices has a nonlocal form.
Moreover, the free field solutions are \cite{Alfaro:2015fha}
\begin{equation}
\psi\left(x\right)=\sum_{r}\int\frac{d^{3}p}{\left(2\pi\right)^{\frac{3}{2}}}\sqrt{\frac{m_{e}}{E_{p}}}\left[b_{r}\left(p\right)u_{r}\left(p\right)e^{-ipx}+d_{r}^{\dagger}\left(p\right)v_{r}\left(p\right)e^{ipx}\right]
\end{equation}
and the fermion dispersion relation $E_{p}=\sqrt{p^{2}+\mu^{2}}$, with the fermion mass $\mu^{2}=m^{2}+m_{e}^{2}$, and
\begin{equation}
A_{\mu}\left(x\right)=\sum_{\lambda}\int\frac{d^{3}k}{\left(2\pi\right)^{\frac{3}{2}}}\sqrt{\frac{1}{2\omega_{k}}}\left[a_{\lambda}\left(k\right)\epsilon_{\mu}^{\lambda}\left(k\right)e^{-ikx}+a_{\lambda}^{\dagger}\left(k\right)\epsilon_{\mu}^{\lambda}\left(-k\right)e^{ikx}\right]
\end{equation}
and the photon dispersion relation $\omega_{k}=\sqrt{k^{2}+m^{2}}$. With these expressions we can evaluate the propagator for the fermionic field
\begin{equation} \label{eq.17a}
S\left(p\right)=i\frac{\gamma.\tilde{p}+m_{e}}{\tilde{p}^{2}-m_{e}^{2}} 
\end{equation}
and also the propagator for the gauge field
\begin{equation}
iD^{\mu\nu}\left(p\right)=\frac{ \eta^{\mu\nu}}{\tilde{k}^{2}},
\end{equation}
in the Feynman gauge $\alpha=1$, observing that the gauge propagator has a massive pole $\tilde{k}^{2} = k^2 -m^2$.
In this case, we see that besides giving massive modes for the fields, VSR appears as nonlocal contributions in the vertices  \eqref{eq.17} and  \eqref{eq.17b}, and propagator  \eqref{eq.17a}.
In particular, the photon propagates massive modes in a gauge invariant framework without changing its number of physical d.o.f. \cite{Bufalo:2019qot,Cheon:2009zx}.

More importantly, in order to evaluate the Feynman amplitude for the Bhabha and Compton scattering processes, we must establish  the energy projection operators in the fermion sector, that are defined as
\begin{equation}
\Pi^{\pm}=\frac{\pm\gamma.\tilde{p}+m_{e}}{2m_{e}}
\end{equation}
so that the completeness relations are written as
\begin{align} \label{eq.10a}
\Pi_{ab}^{+}\left(p\right) & =\sum_{s}u_{a}\left(p,s\right)\overline{u}_{b}\left(p,s\right)=\frac{\gamma.\tilde{p}+m_{e}}{2m_{e}}\\
\Pi_{ab}^{-}\left(p\right) & =-\sum_{s}v_{a}\left(p,s\right)\overline{v}_{b}\left(p,s\right)=\frac{-\gamma.\tilde{p}+m_{e}}{2m_{e}}\label{eq.10b}
\end{align}
the wiggle momentum is $\tilde{p}_\mu = p_\mu -\frac{m^2 n_\mu}{2 (n.p)}$, so that we see the presence of a nonlocal factor in the $\Pi^{\pm}$ energy operators.

For the gauge sector, the completeness relation reads $ \sum_{\lambda} \varepsilon_{\nu}^{*}(\lambda)\varepsilon_{\mu}(\lambda)=-\eta_{\nu\mu}+\text{ gauge terms}$.
Although it is not of our interest, an explicit expression for the polarization vector $\varepsilon_{\mu}(\lambda)$ can be obtained by imposing the wiggled Lorenz condition $\tilde{k}^{\nu}\varepsilon_{\nu}=0$, or alternatively from the complementary conditions $n.\varepsilon=0$ and $k.\varepsilon=0$.

\section{VSR tree-level scattering processes}
\label{sec3}

Scattering of standard model particles is an important scenario where deviations from standard physics can be widely explored, providing a ground and rich framework for testing many theoretical proposals in QFT.
Furthermore, Bhabha scattering, $e^{-}+e^{+}\to e^{-}+e^{+}$, and Compton scattering, $e^{-}+\gamma\to e^{-}+\gamma$, are the most fundamental reactions in QED processes, so that its precision data have been used to obtain bounds in the context of Lorentz violating models \cite{Altschul:2004xp,Charneski:2012py,Casana:2012vu,Bufalo:2015eia,deBrito:2016zav}.

\subsection{Bhabha scattering}

\begin{figure}[t]
\vspace{-1.2cm}
\includegraphics[height=9\baselineskip]{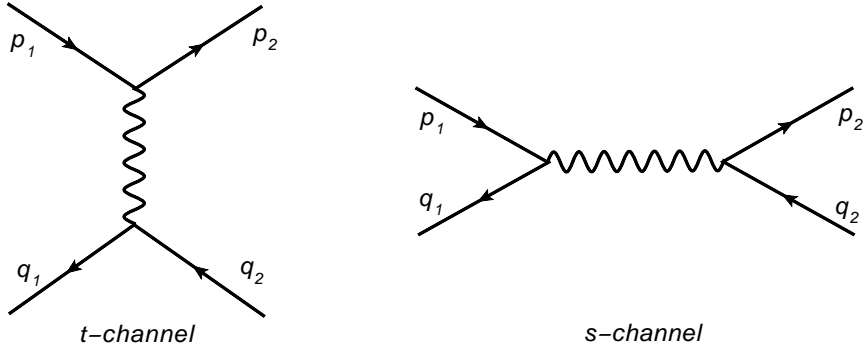}
\centering\caption{Diagrams contributing to the Bhabha scattering: (a) photon direct process and (b) pair annihilation process.}
\label{bhabha}
\end{figure}

We start by considering the Bhabha scattering $e^{-}+e^{+}\to e^{-}+e^{+}$, we have
two possible processes: the t-channel or direct process, and the
s-channel or annihilation process; these are depicted in Fig.~\ref{bhabha}.
We shall consider the standard description for our system:
the spin and momenta for the incoming $u_{r_{i}}\left(p_{1}\right)$ and outgoing $\overline{u}_{r_{f}}\left(p_{2}\right)$ electrons
are $\left(r_i,p_{1}\right)$ and $\left(r_f,p_{2}\right)$, respectively, while the spin and momenta for the incoming $\overline{v}_{s_{i}}\left(q_{1}\right)$ and outgoing $v_{s_{f}}\left(q_{2}\right)$ positrons are $\left(s_i,q_{1}\right)$ and $\left(s_f,q_{2}\right)$, respectively.

The tree-level amplitude corresponds to the sum of the above contributions
\begin{equation}
i\mathfrak{M}=i\mathfrak{M}_{s}+i\mathfrak{M}_{t}
\end{equation}
where the amplitude corresponding to the t-channel reads
\begin{equation}
\mathfrak{M}_{s}=ie^{2}N_{b}\overline{u}_{r_{f}}\left(p_{2}\right)\Lambda^{\mu}\left(p_{1},-q_{1}\right)v_{s_{f}}\left(q_{2}\right)D_{\mu\nu}\left(p_{1}+q_{1}\right)\overline{v}_{s_{i}}\left(q_{1}\right)\Lambda^{\nu}\left(-p_{2},q_{2}\right)u_{r_{i}}\left(p_{1}\right)
\end{equation}
whereas the amplitude for the s-channel is
\begin{equation}
\mathfrak{M}_{t}=-ie^{2}N_{b}\overline{u}_{r_{f}}\left(p_{2}\right)\Lambda^{\mu}\left(p_{1},p_{2}\right)u_{r_{i}}\left(p_{1}\right)D_{\mu\nu}\left(p_{1}-p_{2}\right)\overline{v}_{s_{i}}\left(q_{1}\right)\Lambda^{\nu}\left(q_{1},q_{2}\right)v_{s_{f}}\left(q_{2}\right)
\end{equation}
with the normalization factor defined as
\begin{equation}
N_{b}=\sqrt{\frac{m_{e}}{E_{p_{1}}}}\sqrt{\frac{m_{e}}{E_{p_{2}}}}\sqrt{\frac{m_{e}}{E_{q_{1}}}}\sqrt{\frac{m_{e}}{E_{q_{2}}}}
\end{equation}

Since we are interested in computing the differential cross section for the Bhabha scattering, we shall now compute $\left|i\mathfrak{M}\right|^{2}$ by averaging over the spin of the incoming particles and summing over the spin of the outgoing particles, thus
\begin{equation}
\frac{1}{4}\sum_{r_{i},s_{i}}\sum_{r_{f},s_{f}}\left|i\mathfrak{M}\right|^{2}=\frac{1}{4}\sum_{\textrm{spin}}\left|\mathfrak{M}_{s}\right|^{2}+\frac{1}{4}\sum_{\textrm{spin}}\left|\mathfrak{M}_{t}\right|^{2}+\frac{1}{4}\sum_{\textrm{spin}}\mathfrak{M}_{t}^{*}\mathfrak{M}_{s}+\frac{1}{4}\sum_{\textrm{spin}}\mathfrak{M}_{s}^{*}\mathfrak{M}_{t}
\end{equation}
The sum over spins can be evaluated as usual with help of the completeness relations \eqref{eq.10a} and \eqref{eq.10b}, which yields
\begin{align}
\sum_{\textrm{spin}}\left|\mathfrak{M}_{t}\right|^{2}  =&\frac{e^{4}N_{b}^{2}}{16m_{e}^{4}\left(\tilde{p}_{1}-\tilde{p}_{2}\right)^{4}}\textrm{Tr}\left[\Lambda_{\mu}\left(q_{1},q_{2}\right)\left(\gamma.\tilde{q}_{1}-m_{e}\right)\Lambda_{\alpha}\left(q_{1},q_{2}\right)\left(\gamma.\tilde{q}_{2}-m_{e}\right)\right]\cr
&\times \textrm{Tr}\left[\left(\gamma.\tilde{p}_{1}+m_{e}\right)\Lambda^{\mu}\left(p_{1},p_{2}\right)\left(\gamma.\tilde{p}_{2}+m_{e}\right)\Lambda^{\alpha}\left(p_{1},p_{2}\right)\right]
\end{align}
also
\begin{align}
\sum_{\textrm{spin}}\mathfrak{M}_{s}^{*}\mathfrak{M}_{t}  & =-\frac{e^{4}N_{b}^{2}}{16m_{e}^{4}\left(\tilde{p}_{1}-\tilde{p}_{2}\right)^{2}\left(\tilde{p}_{1}+\tilde{q}_{1}\right)^{2}} \textrm{Tr}\bigg[\Lambda^{\nu}\left(-p_{2},q_{2}\right)\left(\gamma.\tilde{q}_{1}-m_{e}\right)\Lambda_{\mu}\left(q_{1},q_{2}\right)  \cr
&\times \left(\gamma.\tilde{q}_{2}-m_{e}\right)\Lambda_{\nu}\left(p_{1},-q_{1}\right)\left(\gamma.\tilde{p}_{2}+m_{e}\right)\Lambda^{\mu}\left(p_{1},p_{2}\right)\left(\gamma.\tilde{p}_{1}+m_{e}\right)\bigg] .
\end{align}
The remaining two terms can be determined by the substitution $\left(q_{1}\leftrightarrow-p_{2}\right)$ in these two expressions.
In order to highlight the effects from VSR, we shall consider the behavior of the scattering process in the high-energy limit, that corresponds to take $m^2_e =0$ in our calculations.
In this case, we  proceed to the evaluation of the $\gamma$ matrices traces using the standard techniques, and we find that the squared Feynman amplitude reads
\begin{align} \label{eq.11}
\frac{1}{4}\sum_{r_{i},s_{i}}\sum_{r_{f},s_{f}}\left|i\mathfrak{M}\right|^{2} & = \frac{e^{4}}{64E^{4}}\biggl[\frac{\mathcal{J}_1}{\left(\tilde{p}_{1}-\tilde{p}_{2}\right)^{4}}+\frac{\mathcal{J}_2}{\left(\tilde{p}_{1}+\tilde{q}_{1}\right)^{4}}-\frac{ \mathcal{J}_3 }{\left(\tilde{p}_{1}-\tilde{p}_{2}\right)^{2}\left(\tilde{p}_{1}+\tilde{q}_{1}\right)^{2}}\biggr]
\end{align}
where we have defined the following quantities as the annihilation term:
\begin{align}
\mathcal{J}_1  &=32  \left(\tilde{q}_{2}.\tilde{p}_{2}\right)\left(\tilde{q}_{1}.\tilde{p}_{1}\right)+ 32\left(\tilde{p}_{1}.\tilde{q}_{2}\right)\left(\tilde{q}_{1}.\tilde{p}_{2}\right) \nonumber \\
 & +m^{2}\Xi\Big[ \left(\tilde{q}_{2}.\tilde{p}_{2}\right)\left(n.\tilde{q}_{1}\right)\left(n.\tilde{p}_{1}\right)+\left(n.\tilde{q}_{1}\right)\left(n.\tilde{p}_{2}\right)\left(\tilde{q}_{2}.\tilde{p}_{1}\right)-2\left(\tilde{p}_{2}.\tilde{p}_{1}\right)\left(n.\tilde{q}_{2}\right)\left(n.\tilde{q}_{1}\right)\Big] \nonumber \\
 & +m^{2}\Xi\Big[ \left(\tilde{q}_{1}.\tilde{p}_{2}\right)\left(n.\tilde{q}_{2}\right)\left(n.\tilde{p}_{1}\right)+\left(n.\tilde{q}_{2}\right)\left(n.\tilde{p}_{2}\right)\left(\tilde{q}_{1}.\tilde{p}_{1}\right)-2\left(\tilde{q}_{2}.\tilde{q}_{1}\right)\left(n.\tilde{p}_{2}\right)\left(n.\tilde{p}_{1}\right) \Big] \cr
 & +m^{4}\Xi^{2}\left(n.\tilde{q}_{1}\right)\left(n.\tilde{q}_{2}\right)\left(n.\tilde{p}_{2}\right)\left(n.\tilde{p}_{1}\right) 
\end{align}
and the direct term
\begin{align}
\mathcal{J}_2  &=32  \left(\tilde{q}_{2}.\tilde{q}_{1}\right)\left(\tilde{p}_{2}.\tilde{p}_{1}\right)+32\left(\tilde{p}_{1}.\tilde{q}_{2}\right)\left(\tilde{q}_{1}.\tilde{p}_{2}\right)  \nonumber \\
 & -m^{2}\Theta \Big[  \left(\tilde{q}_{2}.\tilde{q}_{1}\right)\left(n.\tilde{p}_{2}\right)\left(n.\tilde{p}_{1}\right)+\left(n.\tilde{q}_{1}\right)\left(n.\tilde{p}_{2}\right)\left(\tilde{q}_{2}.\tilde{p}_{1}\right)-2\left(\tilde{q}_{1}.\tilde{p}_{1}\right)\left(n.\tilde{q}_{2}\right)\left(n.\tilde{p}_{2}\right) \Big] \nonumber \\
 & -m^{2}\Theta \Big[  \left(\tilde{q}_{1}.\tilde{p}_{2}\right)\left(n.\tilde{q}_{2}\right)\left(n.\tilde{p}_{1}\right)+\left(n.\tilde{q}_{2}\right)\left(n.\tilde{q}_{1}\right)\left(\tilde{p}_{2}.\tilde{p}_{1}\right)-2\left(\tilde{q}_{2}.\tilde{p}_{2}\right)\left(n.\tilde{q}_{1}\right)\left(n.\tilde{p}_{1}\right) \Big] \cr
 & +m^{4}\Theta^{2}\left(n.\tilde{q}_{1}\right)\left(n.\tilde{q}_{2}\right)\left(n.\tilde{p}_{2}\right)\left(n.\tilde{p}_{1}\right) 
\end{align}
and finally the interference term
\begin{align}
\mathcal{J}_3  =&-32\left(\tilde{q}_{1}.\tilde{p}_{2}\right)\left(\tilde{q}_{2}.\tilde{p}_{1}\right) -32\left(\tilde{q}_{1}.\tilde{p}_{2}\right)\left(\tilde{q}_{2}.\tilde{p}_{1}\right) \nonumber \\
 & +8m^{2}\left(\Theta+\Xi\right)\Big[ -\left(n.\tilde{p}_{2}\right)\left(n.\tilde{q}_{2}\right)\left(\tilde{q}_{1}.\tilde{p}_{1}\right)+\left(n.\tilde{p}_{1}\right)\left(n.\tilde{p}_{2}\right)\left(\tilde{q}_{1}.\tilde{q}_{2}\right) \cr
 &  -\left(n.\tilde{q}_{1}\right)\left(n.\tilde{p}_{1}\right)\left(\tilde{q}_{2}.\tilde{p}_{2}\right)+\left(n.\tilde{q}_{1}\right)\left(n.\tilde{q}_{2}\right)\left(\tilde{p}_{2}.\tilde{p}_{1}\right)  + \left(n.\tilde{q}_{1}\right)\left(n.\tilde{q}_{2}\right)\left(\tilde{p}_{2}.\tilde{p}_{1}\right)\cr
 & -\left(n.\tilde{p}_{1}\right)\left(n.\tilde{q}_{1}\right)\left(\tilde{p}_{2}.\tilde{q}_{2}\right)+\left(n.\tilde{p}_{2}\right)\left(n.\tilde{p}_{1}\right)\left(\tilde{q}_{2}.\tilde{q}_{1}\right)-\left(n.\tilde{p}_{2}\right)\left(n.\tilde{q}_{2}\right)\left(\tilde{q}_{1}.\tilde{p}_{1}\right)\Big]\nonumber \\
 &-8m^{4}\Theta\Xi\left(n.\tilde{q}_{1}\right)\left(n.\tilde{q}_{2}\right)\left(n.\tilde{p}_{2}\right)\left(n.\tilde{p}_{1}\right)  -8m^{4}\Theta \Xi \left(n.\tilde{q}_{1}\right)\left(n.\tilde{q}_{2}\right)\left(n.\tilde{p}_{2}\right)\left(n.\tilde{p}_{1}\right)
\end{align}
and by simplicity we have introduced the notation
\begin{equation}
\Xi=\frac{1}{\left(n.p_{1}\right)\left(n.p_{2}\right)}+\frac{1}{\left(n.q_{1}\right)\left(n.q_{2}\right)}, \quad
\Theta=\frac{1}{\left(n.p_{1}\right)\left(n.q_{1}\right)}+\frac{1}{\left(n.p_{2}\right)\left(n.q_{2}\right)}.
\end{equation}

Now to cast the expression \eqref{eq.11} in a suitable form for evaluation, we consider the kinematic variables in the center-of-mass (CM) frame, so that we have
\begin{equation}
\begin{cases}
p_{1}=\left(E,\vec{p}\right)\\
p_{2}=\left(E,-\vec{p}\right)
\end{cases}\begin{cases}
q_{1}=\left(E,\vec{q}\right)\\
q_{2}=\left(E,-\vec{q}\right)
\end{cases}
\end{equation}
It follows from the energy conservation that $|\vec{p}| = |\vec{q}|$, and we also introduce $\theta$ as the center-of-mass scattering angle, i.e. $(\vec{p}.\vec{q}) = p^2 \cos\theta$.
Finally, we can proceed to the computation of the differential cross section (in natural units) that in the CM frame is given by
\begin{equation}
\left(\frac{d\sigma}{d\Omega }\right)_{\textrm{cm}}  =\frac{E^{2}}{16\pi^{2}} \frac{1}{4}\sum_{r_{i},s_{i}}\sum_{r_{f},s_{f}}\left|i\mathfrak{M}\right|^{2}
\end{equation}
Hence, using the result \eqref{eq.11} in terms of the CM variables we can express the cross section as follows
\begin{align}
\left(\frac{d\sigma}{d\Omega}\right)_{\textrm{cm}} & =\frac{\alpha^{2}}{8E^{2}}\biggl[\frac{1}{2}\left(1-\chi^{2}\right)^{2}\left(1+\cos^{2}\theta\right)-\frac{1}{16}\chi^{2}\left(1-\frac{3}{2}\chi^{2}\right)\nonumber \\
 & +\frac{1}{\left(\sin^{2}\frac{\theta}{2}+\frac{\chi^{2}}{2}\cos\theta\right)^{2}}\left(\left(1-\chi^{2}\right)^{2}\left[1+\cos^{4}\frac{\theta}{2}\right]-\frac{1}{32}\chi^{2}\left(1-\chi^{2}\right)\left( 1+3\cos\theta\right) +\frac{1}{32}\chi^{4}\right)\nonumber \\
 & -\frac{1}{\left(\sin^{2}\frac{\theta}{2}+\frac{\chi^{2}}{2}\cos\theta\right)}\left(2\left(1-\chi^{2}\right)^{2}\cos^{4}\frac{\theta}{2}-\chi^{2}\left(1-\frac{3}{2}\chi^{2}\right)\right)\biggr]
\end{align}
where we have introduced the parameter $\chi=\frac{m}{E}$, which controls perturbatively the VSR effects. Finally, in order to illustrate the corrections due to the VSR nonlocal effects, we express the differential cross section up to $\chi^2$ order, resulting into
\begin{align} \label{eq.12}
\left(\frac{d\sigma}{d\Omega}\right)_{\textrm{cm}} & =\frac{\alpha^{2}}{8E^{2}}\left(\frac{1+\cos^{4}\frac{\theta}{2}}{\sin^{4}\frac{\theta}{2}}-\frac{2\cos^{4}\frac{\theta}{2}}{\sin^{2}\frac{\theta}{2}}+\frac{1+\cos^{2}\theta}{2}\right) \cr
 & +\frac{\alpha^{2}}{8E^{2}}\chi^{2}\bigg(-\frac{\left(1+\cos^{4}\frac{\theta}{2}\right)\cos\theta}{\sin^{6}\frac{\theta}{2}}+\frac{32\cos^{4}\frac{\theta}{2}\left(\cos\theta-2\right)-3\cos\theta-65}{32\sin^{4}\frac{\theta}{2}} \cr
 &+\frac{1+4\cos^{4}\frac{\theta}{2}}{\sin^{2}\frac{\theta}{2}}-\cos^{2}\theta-\frac{17}{16}\bigg)
\end{align}
where we identify the first term as the usual QED contribution to the Bhabha scattering, while the second term is solely due to VSR.
It is worth noticing that this VSR correction term has a similar structure as the QED one, since it has the same energy behavior as  $1/E^2$, and depends only on the scattering angle $\theta$. 
In addition to the result found to the Bhabha process \eqref{eq.12}, we could use crossing symmetry arguments to also obtain the VSR modified differential cross section to the M\o ller process, that consists in the electron-electron scattering $e^-+e^- \to e^- +e^-  $.

To conclude the discussion of this section, it is worth to recall that the study of Bhabha scattering is still relevant mainly because it is the process employed in determining the luminosity $L$ at $e^+ e^-$ colliders \cite{Barate:1999ce}.
In particular, there are two kinematical regions of interest for the luminosity measurements: one is known as the small-angle Bhabha (SABh) process, which is found at scattering angles below $6^{\circ}$, and is mainly dominated by the t-channel (direct); the other is the large angle Bhabha (LABh) process, which is found at scattering angles above $6^{\circ}$, and receives important contributions from various s-channel (annihilation).
We can assess the VSR correction to Bhabha scattering using the formula
\begin{equation}
\delta  =\left( \frac{d\sigma }{d\Omega }\right)^{\rm VSR}  \left/ \left( \frac{d\sigma }{d\Omega }\right)\right. ^{\rm QED}-1,
\end{equation}
Hence,  we can calculate the leading effects of the VSR approach in the SABh process, by expanding Eq.~\eqref{eq.12} for small angles $\theta \ll 1$, and find that the correction is
\begin{equation}
\delta_{\rm SABh}^{\rm vsr}  = -16\left(\frac{m}{\sqrt{s}}\right)^{2}\frac{1}{\theta^{2}}
\end{equation}
where we have introduced by convenience the usual Mandelstan variable $s = (p_1+q_1)^2$ or in the CM frame $\sqrt{s}=2E$ is the center-of-mass energy.

Furthermore, for the LABh process, we see from \eqref{eq.12} that the VSR correction increases as the angle increases, taking its maximum value at $\theta = 90^{\circ}$. 
Hence, in this case, we have that
\begin{equation}
\delta_{\rm LABh}^{\rm vsr}  = -\frac{49}{6}\left(\frac{m}{\sqrt{s}}\right)^{2}
\end{equation}

These expressions for the VSR deviations $\delta^{\rm vsr}$ can be discussed in high-precision luminosity measurements.
As illustration, either the SABh process the energy corresponding to the Z resonance, at $\sqrt{s}=91\,{\rm GeV}$ \cite{ILC}, or the LABh regime, at $\sqrt{s}=10 \,{\rm GeV}$ \cite{CarloniCalame:2000pz}, gives the same roughly estimate value for the VSR parameter as $m \leq 10^{9}\,{\rm eV}$.
Actually, this bound is not significant if we associate the value of $m$ for instance with the photon mass $m_\gamma \leq 10^{-18}{\rm eV}$ \cite{Tanabashi:2018oca}.
In this sense, although the nonlocal effects from VSR give a wealth departure from the usual QED behavior in the description of the Bhabha scattering, its experimental data do not set strong bounds upon the VSR parameter.

\subsection{Compton scattering}

\begin{figure}[t]

\includegraphics[height=5.5\baselineskip]{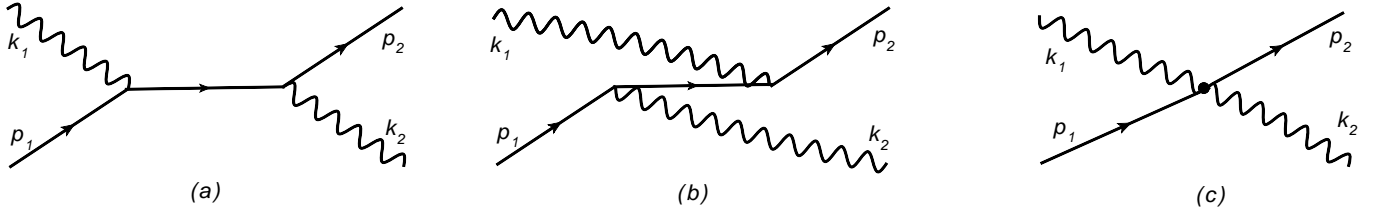}
\centering\caption{Diagrams contributing to the Compton scattering.}
\label{compton}
\end{figure}

Another important process in QED is the Compton scattering $e^{-}+\gamma\to e^{-}+\gamma$ and have been used as probe to study departures from Lorentz symmetry.
The usual expression for the differential cross section cannot be used since the kinematics is modified, i.e., the Compton energy shift (energy-momentum conservation) is modified by VSR effects, and we shall approach this with further detail.

We start by analyzing the kinematics of the Compton process.
It is worth to evaluate this in the laboratory frame where the electron is at rest initially, $p_1=\left(E_{p},\vec{0}\right)$.
Assuming that $p_1+k_1\to p_2+k_2$, we see that the energy-momentum conservation $p_2^{2}=\mu^{2}$ for the outgoing electron yields
\begin{equation} \label{eq.15}
2m^{2}+2E_{p}\left(\omega_{k}-\omega_{k}'\right)-2\left(\omega_{k}\omega_{k}'-\sqrt{\omega_{k}^{2}-m^{2}}\sqrt{ (\omega'_{k})^{2}-m^{2}}\cos\theta\right)=0
\end{equation}
where we have used that $k_1 = (\omega_k, \vec{k}_1)$ and $k_2 = (\omega'_k, \vec{k}_2)$.
We can solve the above expression for $\omega_{k}'$, so that
\begin{align} \label{eq.13}
\gamma  =\frac{\left(\sqrt{1+\kappa^{2}}+\xi\right)\left(\sqrt{1+\kappa^{2}}\xi+\kappa^{2}\right)+\left(\xi^{2}-\kappa^{2}\right)\cos\theta\sqrt{ 1+\kappa^{2}\cos^{2}\theta }}{\left(\sqrt{1+\kappa^{2}}+\xi\right)^{2}-\left(\xi^{2}-\kappa^{2}\right)\cos^{2}\theta}
\end{align}
where we have introduced the following parameters:
\begin{equation} \label{eq.21}
\kappa=\frac{m}{m_{e}},\quad  \xi=\frac{\omega_{k}}{m_{e}},\quad  \gamma=\frac{\omega_{k}'}{m_{e}}
\end{equation}
Notice that $\kappa=m/m_{e}$ controls the VSR modifications in the expression \eqref{eq.13}. In particular, observe that in the limit $\kappa^2\to0$ equation \eqref{eq.13} reproduces the known Compton relation
\begin{equation} \label{eq.14}
\omega_{k}'=\frac{\omega_{k}}{1+\frac{\omega_{k}}{m_e}\left(1-\cos\theta\right)}
\end{equation}

As previously mentioned, in order to the compute the cross section associated with the Compton scattering, it is necessary to consider the energy shift relation \eqref{eq.13} rather than equation \eqref{eq.14}.
To illustrate the main aspects involving the VSR modification into the differential cross section, we shall consider the main definition
\begin{align} \label{eq.31}
d\sigma & =\frac{\left(2\pi\right)^{4}}{\left(2\pi\right)^{6}2\omega_{k}}\int\delta\left(p_{1}+k_{1}-p_{2}-k_{2}\right)\left|i\mathfrak{M}\right|^{2}\frac{m_{e}d^{3}p_{2}}{E_{p_{2}}}\frac{d^{3}k_{2}}{2\omega_{k_{2}}}
\end{align}
which after some manipulations is written as
\begin{align}  \label{eq.16}
\frac{d\sigma}{d\Omega_{k}'} & =\frac{m_{e}}{8\pi^{2}\omega_{k}}\int d\omega'_{k} \sqrt{(\omega'_{k})^{2}-m^{2}} \left|i\mathfrak{M}\right|^{2}\delta\left(\left(p_{1}+k_{1}-k_{2}\right)^{2}-\mu^{2}\right)\Theta\left(E_{p}+\omega_{k}-\omega_{k}'\right) 
\end{align}
Now in order to solve the integral over the outgoing photon energy $\omega'_{k}$ we must take into account the energy-momentum conservation \eqref{eq.15} that is present in the argument of the $\delta$ function in \eqref{eq.16}.
This integration can be evaluated with help of the identity
\begin{equation}
\int dx\delta\left(f\left(x\right)\right)g\left(x\right)=\sum_{i}\left.\frac{g\left(x_{i}\right)}{\left|\frac{df}{dx}\right|}\right|_{\textrm{zeros of }f\left(x\right)}
\end{equation}
with the identification of $f(\omega'_{k})$ with the equation \eqref{eq.15} for the energy-momentum conservation.
Hence, after applying the above identity and some manipulation, we find the following expression for the VSR modified differential cross section for the Compton scattering:
\begin{align} \label{eq.22}
\left(\frac{d\sigma}{d\Omega_{k}'} \right)_{\rm vsr}  =\frac{1}{16\pi^{2}\xi}\frac{\sqrt{\gamma^{2}-\kappa^{2}} \left|i\mathfrak{M}\right|^{2}}{\sqrt{1+\kappa^{2}}+\xi\left(1-\cos\theta\right)-\frac{\kappa^{2}}{\sqrt{1+\kappa^{2}}}\cos\theta\left(1-\cos\theta\right)\left[ 1+\frac{\xi}{2\sqrt{1+\kappa^{2}}}\left(1-\cos\theta\right)\right] }
\end{align}
observe that in the limit $\kappa^2 \to 0$ this expression reduces to
\begin{align}
\left(\frac{d\sigma}{d\Omega_{k}'} \right)_{\rm qed}  =\frac{1}{16\pi^{2}} \left( \frac{\omega'_k}{\omega_k}\right)^2  \left|i\mathfrak{M}\right|^{2}
\end{align}
reproducing the known QED result, where $\omega'_k$ is given by \eqref{eq.14}.

With help of the Feynman rules for the vertex function and propagators, we can proceed and compute the amplitude related to the graphs, Fig.~\ref{compton}, which results into
\begin{align}
i\mathfrak{M} & =i\mathfrak{M}_{a}+i\mathfrak{M}_{b}+i\mathfrak{M}_{c} \cr
 & \equiv i e^{2} \overline{u}_{r_{f}}\left(p_{2}\right)\Gamma^{\alpha\beta}u_{r_{i}}\left(p_{1}\right)\epsilon_{\alpha}\left(k_{1},\lambda\right)\epsilon_{\beta}\left(k_{2},\lambda'\right)
\end{align}
where we have defined by simplicity the quantity
\begin{align} 
\Gamma^{\alpha\beta} & =\gamma^{\beta}\frac{\gamma.\left(\tilde{p}_{1}+\tilde{k}_{1}\right)+m_{e}}{\left(\tilde{p}_{1}+\tilde{k}_{1}\right)^{2}-m_{e}^{2}}\gamma^{\alpha}+\gamma^{\alpha}\frac{\gamma.\left(\tilde{p}_{1}-\tilde{k}_{2}\right)+m_{e}}{\left(\tilde{p}_{1}-\tilde{k}_{2}\right)^{2}-m_{e}^{2}}\gamma^{\beta} \cr
&+  \frac{ m^2}{2} \frac{(n.\gamma) n^\alpha n^\beta }{(n.k_1)(n.k_2)} \left[  \frac{1}{\left(n.p_{1}\right)} + \frac{1}{\left(n.p_{2}\right)} - \frac{1}{n.\left(p_{1}+k_1\right)}- \frac{1}{n.\left(p_{1}-k_2\right)}   \right].\label{eq.18a}
\end{align}
It is easy to see that the ``seagull'' term $i\mathfrak{M}_{c} $ in the expression \eqref{eq.18a} vanishes identically under the VSR gauge condition ($n.\varepsilon=0$ and $k.\varepsilon=0$), showing that the quartic vertex \eqref{eq.17b} does not contribute to the Compton scattering. Hence, we have that
\begin{align} 
\Gamma^{\alpha\beta}  =\gamma^{\beta}\frac{\gamma.\left(\tilde{p}_{1}+\tilde{k}_{1}\right)+m_{e}}{\left(\tilde{p}_{1}+\tilde{k}_{1}\right)^{2}-m_{e}^{2}}\gamma^{\alpha}+\gamma^{\alpha}\frac{\gamma.\left(\tilde{p}_{1}-\tilde{k}_{2}\right)+m_{e}}{\left(\tilde{p}_{1}-\tilde{k}_{2}\right)^{2}-m_{e}^{2}}\gamma^{\beta} .\label{eq.18}
\end{align}

Since we are interested to evaluate the cross section for the Compton scattering for unpolarized photons, in this case we shall compute $\left|i\mathfrak{M}\right|^{2}$ by taking an average over the initial spin and polarization and sum over the
final spin and polarization,
\begin{align} \label{eq.19}
\frac{1}{4}\sum_{r_{i},r_{f}}\sum_{\lambda,\lambda'}\left|i\mathfrak{M}\right|^{2}  =\frac{e^{4} }{16m_{e}^{2}}\textrm{Tr}\left[\Gamma_{\alpha\beta}^{\dagger}\left(\gamma.\tilde{p}_{2}+m_{e}\right)\Gamma^{\alpha\beta}\left(\gamma.\tilde{p}_{1}+m_{e}\right)\right]
\end{align}
where completeness relations for the fermion \eqref{eq.10a} and for the gauge field were used.
After some algebraic simplification, we can compute the trace of the $\gamma$ matrices.
We can also make use of the scattering kinematics variables and cast the expression: \eqref{eq.19} as the following
\begin{align} \label{eq.20}
\frac{1}{4}\sum_{r_{i},r_{f}}\sum_{\lambda,\lambda'}\left|i\mathfrak{M}\right|^{2} & =\frac{e^{4} }{2m_{e}^{2}}\frac{1}{\left(\tilde{p}_{1}.\tilde{k}_{1}\right)}\frac{1}{\left(\tilde{p}_{1}.\tilde{k}_{1}\right)}\left[m_{e}^{4}+m_{e}^{2}\left(\tilde{p}_{1}.\tilde{k}_{1}\right)+\left(\tilde{p}_{1}.\tilde{k}_{1}\right)\left(\tilde{p}_{1}.\tilde{k}_{2}\right)\right]\nonumber \\
 & -\frac{e^{4} }{2m_{e}^{2}}\frac{1}{\left(\tilde{p}_{1}.\tilde{k}_{2}\right)}\frac{1}{\left(\tilde{p}_{1}.\tilde{k}_{1}\right)}\left[2m_{e}^{4}+m_{e}^{2}\left(\tilde{p}_{1}.\tilde{k}_{1}\right)-m_{e}^{2}\left(\tilde{p}_{1}.\tilde{k}_{2}\right)\right]\nonumber \\
 & +\frac{e^{4} }{2m_{e}^{2}}\frac{1}{\left(\tilde{p}_{1}.\tilde{k}_{2}\right)}\frac{1}{\left(\tilde{p}_{1}.\tilde{k}_{2}\right)}\left[m_{e}^{4}-m_{e}^{2}\left(\tilde{p}_{1}.\tilde{k}_{2}\right)+\left(\tilde{p}_{1}.\tilde{k}_{1}\right)\left(\tilde{p}_{1}.\tilde{k}_{2}\right)\right]
\end{align}

Remember that the expression \eqref{eq.22} for the differential cross section was computed in the laboratory frame; hence, a necessary last step before its evaluation is to express equation \eqref{eq.20} in terms of the laboratory frame variables.
In this case, we have that $p_1=\left(\mu,\vec{0}\right)$, and using the definition $\tilde{p}_\mu = p_\mu - \frac{m^2}{2} \frac{n_\mu}{(n.p)}$, we find that
\begin{align} 
\left(\tilde{p}_{1}.\tilde{k}_{1}\right) &=  m_{e}^{2}\left[\sqrt{1+\kappa^{2}}\xi-\frac{\kappa^{2}\sqrt{1+\kappa^{2}}}{2\left(\xi-\sqrt{\xi^{2}-\kappa^{2}}\right)}-\frac{\kappa^{2}\left(\xi-\sqrt{\xi^{2}-\kappa^{2}}\right)}{2\sqrt{1+\kappa^{2}}}\right] \label{eq.24a}\\ 
\left(\tilde{p}_{1}.\tilde{k}_{2}\right) &= m_{e}^{2}\left[\sqrt{1+\kappa^{2}}\gamma-\frac{\kappa^{2}\sqrt{1+\kappa^{2}}}{2\left(\gamma-\sqrt{\gamma^{2}-\kappa^{2}}\cos\theta\right)}-\frac{\kappa^{2}\left(\gamma-\sqrt{\gamma^{2}-\kappa^{2}}\cos\theta\right)}{2\sqrt{1+\kappa^{2}}}\right] \label{eq.24}
\end{align}
where we have used the set of parameters defined in \eqref{eq.21} by the simplicity of notation.
Notice that the polar angle $\theta$ is present in \eqref{eq.20} in two forms: in the definition of the energy-shift relation $\gamma$ in Eq.~\eqref{eq.13}, but also by means of the VSR anisotropy in the term $\left(\tilde{p}_{1}.\tilde{k}_{2}\right)$ expressed in Eq.~\eqref{eq.24}.

Now that we have fully determined the squared Feynman amplitude in terms of the laboratory variables, obtained by replacing Eqs.~\eqref{eq.24a} and  \eqref{eq.24} back into \eqref{eq.20}, we can proceed to the evaluation of the cross section by means of the integration $\sigma   =\int\sin\theta d\theta d\varphi\frac{d\sigma}{d\Omega_{k}'}$.

Let us now consider some particular cases, where the integration can
be easily solved.
\begin{enumerate}

\item For the case of small photons energies $\xi\ll1$, we can
determine the leading VSR contributions in $\kappa^2$ to the total cross section as follows:
\begin{equation}
\left. \sigma \right|_{\xi\ll1} =\sigma_{\textrm{Th}}+\delta\sigma_{\textrm{vsr}}
\end{equation}
where $\sigma_{\textrm{Th}}=\frac{8\pi}{3}\frac{\alpha^{2}}{m_{e}^{2}}$
is the known classical Thomson cross section and $\alpha= e^2/4\pi$ is the fine-structure constant, while the VSR correction reads
\begin{equation} \label{eq.100}
\delta\sigma_{\textrm{vsr}}=-\frac{\pi\alpha^{2}}{m_{e}^{2}}\left[\frac{117}{5}-\frac{8}{\xi}+\frac{8}{3\xi^{2}}-\frac{1}{\xi^{3}}\left(2-\ln2\right)\right]\kappa^{2}
\end{equation}
One can observe that the main difference, besides the presence of the minus sign, which signals a reduction of the value of the cross section, is that the VSR contribution has terms that depend on the incoming photon energy $\xi=\omega_k/m_{e}$.

\item For the case of high photons energies $\xi\gg1$, we obtain that the leading VSR correction to the total cross section is
\begin{align} \label{eq.101}
\left. \sigma \right|_{\xi\gg 1}  & =\frac{\pi\alpha^{2}}{m_{e}^{2}\xi}\left[\ln2\xi-\frac{1}{2}+\kappa^{2}\left(3-\ln2\xi-\frac{3\xi}{4}\right)\right]
\end{align}
Notice that although the VSR contribution has different energy dependence from the usual QED term, it respects the point of view that the VSR expansion is seen as a correction and decreases the value of the total cross section.

\item At last, we can present the exact integration for any values of $\xi$,
corresponding to a generalization of the Klein-Nishina cross section,
\begin{align}
\sigma &=\frac{\pi\alpha^{2}}{m_{e}^{2}}\bigg[\frac{2\xi\left(\xi+1\right)\left(\xi+8\right)+4}{\xi^{2}\left(2\xi+1\right)^{2}}+\frac{\left(\xi^{2}-2\xi-2\right)\ln\left(2\xi+1\right)}{\xi^{3}} \\
&+\frac{\kappa^{2}}{2\xi^{4}} \left(\frac{4+24\xi+ 42\xi^{2} +6\xi^{3} -46\xi^{4} -16\xi^{5}+16\xi^{6}}{\left(1+2\xi\right)^{3}}- \frac{\left(2\xi^{4}-2\xi^{3}-3\xi^{2}+2\right)\ln\left(1+2\xi\right)}{\xi}\right)\bigg] \nonumber
\end{align}

\end{enumerate}

As a last remark, it is important to observe that the obtained VSR departures from the QED results come from nonlocal contributions present in the propagators and vertex function; the main interesting aspects of these effects are the departure of the usual expression and the novel energy dependence seen in Eqs.~\eqref{eq.100} and \eqref{eq.101}.

%%%%%%%%%%%%%%%%%%%%%%%%%%%%%%%%%%%%%%%%%%%%%%%%%%%%%%%%%%%%%%%%%%%%%%%
%%%%%%%%%%%%%%%%%%%%%%%%%%%%%%%%%%%%%%%%%%%%%%%%%%%%%%%%%%%%%%%%%%%%%%%
\section{Final remarks}
\label{conc}

In this paper, we presented a study on the Bhabha and Compton scattering processes of the quantum electrodynamics in the VSR Lorentz violating framework.
Our main interest in the paper was to establish the leading behavior of the VSR effects into the differential cross section to the given processes.
For that purpose, we first revised the main aspects regarding the VSR quantum electrodynamics, by establishing the respective Feynman rules, free field solutions, energy projection operators and the completeness relations for the fermion and gauge fields.
It is worth to emphasize that the VSR modified gauge invariance is important not only to the presence of new couplings, but it also allows the photon to propagate massive modes in a gauge invariant framework without changing its number of physical d.o.f. \cite{Bufalo:2019qot,Cheon:2009zx}, i.e. the VSR massive photon has two polarization states, not presenting a third and longitudinal state as in the Proca's massive electrodynamics.

We started by investigating the VSR effects into the cross section of the Bhabha scattering. 
The definition for the differential cross section in the VSR framework holds as usual, so our main work was to evaluate the corresponding Feynman amplitude for the two contributions.
In order to evaluate the cross-section, we expressed the kinematic variables in the CM frame, where the calculation and physical significance become clearer.
After proceeding with a detailed computation, 
It is worth to remark that the VSR effects have the same energy profile $1/E^2$ of the QED, and depends mainly on the scattering angle $\theta$, showing therefore that the VSR modification is minimal into the cross section.
In conclusion, we presented a discussion on the leading VSR modifications in the differential cross section in the case of the small- and large-angle Bhabha processes.

In the discussion of the Compton process in the VSR framework, caution is needed in the computation of the cross section, mainly because VSR modifies the process kinematics.
Hence, in order to have a differential cross section consistent with the VSR effects, we started from the very definition in terms of the final state variables  \eqref{eq.31}, and after some manipulations with the consideration of the VSR energy-momentum conservation, we found a VSR modified cross section \eqref{eq.22} (defined in the laboratory frame).
After some laborious algebraic evaluation of the Feynman amplitude of the Compton scattering for the case of unpolarized photons, we managed to obtain the VSR Compton differential cross section and proceeded to the discussion of some particular cases in order to highlight the behavior of the leading VSR effects:
(i) we first considered the limit of low-energy photons, $\xi = \omega_k/m_e \ll 1$, where we could establish the VSR correction to the classical Thomson cross section; (ii) in the case of high-energetic photons, $\xi \gg 1$, we found a constant VSR correction to the logarithm energy dependence.

We believe that with this study we have completed a series of previous works related with the VSR modifications of the most important aspects of QED: scattering by an external field and bremsstrahlung \cite{Alfaro:2019koq}, one-loop corrections of QED \cite{Alfaro:2017umk}, and the electroweak theory \cite{Dunn:2006xk,Alfaro:2015fha}.
Certainly further studies involving phenomenological interesting scenarios, for instance, neutrino physics, are necessary in order to elucidate the physical aspects of VSR field theories.

%%%%%%%%%%%%%%%%%%%%%%%%%%%%%%%%%%%%%%%%%%%%%%%%%%%%%%%%%%%%%%%%%%%%%%%
%%%%%%%%%%%%%%%%%%%%%%%%%%%%%%%%%%%%%%%%%%%%%%%%%%%%%%%%%%%%%%%%%%%%%%%
 \subsection*{Acknowledgements}

The authors would like to thank the anonymous referee
for his/her comments and suggestions to improve this
paper.
R.B. acknowledges partial support from Conselho
Nacional de Desenvolvimento Cient\'ifico e Tecnol\'ogico (CNPq Projects No. 304241/2016-4 and No. 421886/2018-8) and Funda\c{c}\~ao de
Amparo \`a Pesquisa do Estado de Minas Gerais (FAPEMIG Project No. APQ-01142-17).

%%%%%%%%%%%%%%%%%%%%%%%%%%%%%%%%%%%%%%%%%%%%%%%%%%%%%%%%%%%%%%%%%%%%%%%%%%%%%%%%%%%%%%%%%%%%%%%%%%%%%%%%%%%%%%%%%%%%%%%%%%%%%%%%%%%%%%%%%%%%%%%%%%%%%%%%%%%%%%%%%%%%%%%%%%%%%%%%%%%%%%%%%%%%%%%%%%%%%%%%%%%%%%%%%%%%%%

\global\long\def\link#1#2{\href{http://eudml.org/#1}{#2}}
 \global\long\def\doi#1#2{\href{http://dx.doi.org/#1}{#2}}
 \global\long\def\arXiv#1#2{\href{http://arxiv.org/abs/#1}{arXiv:#1 [#2]}}
 \global\long\def\arXivOld#1{\href{http://arxiv.org/abs/#1}{arXiv:#1}}

%%%%%%%%%%%%%%%%%%%%%%%%%%%%%%%%%%%%%%%%%%%%%%%%%%%%%%%%%%%%%%%%%%%%%%%%%%%%%%%%%%%%%%%%%%%%%%%%%%%%%%%%%%%

\end{document}